\documentclass{epl}

\title{Macroscopic quantum effects in nanomechanical systems}
\author{P. Werner\inst{1} \and W. Zwerger\inst{1,2}}
\institute{
\inst{1} 
Center for NanoScience and Sektion Physik, 
Ludwig-Maximilians-Universit{\"a}t,\\  
Theresienstr. 37, D- 80333 M{\"u}nchen, Germany.\\
\inst{2}
Institute for Theoretical Physics, University of Innsbruck, Technikerstr. 25,\\
A-6020 Innsbruck, Austria.
}

\pacs{03.65.-w}{Quantum mechanics, coherence, tunnelling}
\pacs{62.25.+g}{Mechanical properties of nanoscale materials}
\pacs{62.30.+d}{Mechanical and elastic vibrations }
\begin{document}

\maketitle

\begin{abstract}
We investigate quantum effects in the mechanical properties of
elastic beams on the nanoscale.
Transverse quantum and thermal fluctuations and the nonlinear 
excitation energies are calculated for beams compressed
in longitudinal direction.  
Near the Euler instability, the system is described by a
one dimensional Ginzburg-Landau model where the order parameter
is the amplitude of the buckling mode.
We show that in single wall carbon nanotubes 
with lengths of order or smaller than 100 nm
zero point fluctuations are accessible
and discuss the possibility of observing macroscopic quantum
coherence in nanobeams near the critical strain.
\end{abstract}

\section{Introduction}
The progress in miniaturization of electromechanical 
devices towards the nanometer scale (NEMS)
is beginning to reach the limit, where quantum effects play an important 
role~\cite{Roukes,Craighead,Cleland}.
For example, in nanoscale beams phonons may propagate ballistically,
leading to a quantized thermal conductance~\cite{Schwab}.
Moreover a sizeable contribution to the forces between plates and 
beams which are separated by less than one micron is the Casimir force 
between neutral objects due to the modification of the electromagnetic vacuum
~\cite{Chan,Buks1}. The combination of electrical and mechanical properties 
may be studied via quantized transverse deflection due to charge quantization
of charged, suspended beams in an electric field~\cite{Zant}.
Similarly the standard Coulomb-blockade
in small metallic islands or in semiconducting quantum dots
may be used to mechanically transfer single electrons with a 
nanomechanical oscillator~\cite{Erbe,Isacsson}.
Regarding possible applications of nanomechanical sensors,
Si-based resonators in the radio-frequency regime
were recently fabricated and manipulated~\cite{Pescini}.
In the present work we focus on quantum 
effects in mechanical resonators on the nanometer scale,
in particular in single wall carbon nanotubes (SWNT).
Due to their small masses and remarkable elastic properties
down to nanometer scale, carbon nanotubes are ideally 
suited to study effects like phonon quantization~\cite{Roukes2},
the generation of non-classical states of mechanical motion
~\cite{Zoller} or
macroscopic quantum tunnelling out of a metastable configuration~\cite{Buks2}.
On the classical level, both the thermal
Brownian motion of single nanotubes clamped on one side~\cite{Treacy}
and the discrete eigenmodes of charged multiwall nanotubes excited 
by an ac-voltage~\cite{Poncharal} have been detected experimentally.
More recently, the thermal vibrations of doubly clamped SWNT's
down to lengths of around $0.5{\rm{\mu m}}$ have been observed with a scanning 
electron microscope~\cite{Schoenenberger}.
In all of these cases it turns out that the measured
transverse vibrations of nanotubes agree reasonably well with the
predictions of an elastic continuum model.
Its applicability even on the nm scale is also supported by
molecular dynamics simulations which show
that SWNT's down to lengths of around $10{\rm{nm}}$ are well described by an 
effective elastic continuum,
responding in a reversible manner up to large deformations 
~\cite{Yakobson}.
In the following, we will therefore use the standard theory of an elastic 
continuum~\cite{Landau} for carbon nanotubes which are clamped
between two fixed end points.
We calculate both thermal and quantum fluctuations 
of the nanotube under longitudinal compression, including properly 
the nonlinearity in the bending energy.
It is shown that in SWNT's with a length below $100{\rm{nm}}$ the crossover 
from thermal to quantum zero point fluctuations is 
reached at accessible temperatures of around $30{\rm{mK}}$.
We also discuss the possibility to realize coherent superpositions
of macroscopically distinct states by observing the avoided 
level crossing near the degenerate situation above the 
critical force of the well known Euler-buckling instability.

\section{The Model}

Our model system is a freely suspended SWNT of length $L$ and diameter $D$ 
which is fixed at both ends, allowing only transverse vibrations.
In addition we consider a mechanical force $F$ which acts
on the beam in longitudinal direction ($F>0$ for compression).
In a classical description the beam  
is then completely described by the transverse deflection $\phi(s)$ 
parametrized by the arclength $s \in [0,L]$.
We assume the beam to be incompressible in longitudinal direction
and only keep a single transverse degree of freedom for simplicity
(see below). 
For arbitrary strong deflections $\phi(s)$
the nonlinear Lagrangian of the system is then~\cite{Poston}
\begin{equation}
\label{eq:Lagrangian}
{\mathcal{L}}=\int_0^L ds \left[
\frac{\sigma}{2} \dot{\phi}^2- 
\frac{\mu}{2} \frac{(\phi'')^2}{(1-(\phi')^2)} -
F(\sqrt{1-(\phi')^2}-1) \right].
\end{equation} 
Here $\sigma=m/L$ is the mass density, while the bending rigidity
$\mu=EI$ is the product of the elasticity modulus $E$ and 
the moment of inertia $I=\pi D^3 d/8$, with $d$ an effective wall thickness.
For small deformations $|\phi'(s)| \ll 1$ the Lagrangian is quadratic,
leading to the standard linear equation of motion
\begin{equation}
\label{eq:eqofmo}
\sigma \ddot{\phi}+\mu \phi''''+F\phi''=0
\end{equation}
for the transverse vibrations of an elastic beam under compression.
The corresponding eigenmodes $\phi_n$ and eigenfrequencies
$\omega_n$ depend on the boundary conditions.
We assume that the experimental realization~\cite{Schoenenberger} 
is well described by
clamped ends at both sides, $\phi(0)=\phi(L)=0$ and $\phi'(0)=\phi'(L)=0$. 
The exact $\phi_n$'s are then given by a superposition of
trigonometric and hyperbolic functions, and the $\omega_n$'s
by solving a transcendental equation~\cite{Landau}.
In the following, for some of the analytic expressions, we will
use boundary conditions without bending moments at the ends
of the beam, $\phi''(0)=\phi''(L)=0$.
This leaves the essential physics unchanged and permits one to
write down simple expressions for the 
eigenfunctions in the normal mode expansion
\begin{equation}
\label{eq:eigenm}
\phi(s) =\sum_n {\mathcal{A}}_n \sin{ \frac{n\pi}{L}s}
\end{equation}
and its eigenfrequencies
\begin{equation}
\label{eq:eigenfr}
\omega_n= \left( \frac{\mu (n\pi/L)^2-F}{\sigma} \right) ^\frac{1}{2} 
\frac{n\pi}{L}.
\end{equation}
Clearly the modes soften with increasing compression $F$,
up to a critical force $F_{c}=\mu (\pi/L)^2$ where the fundamental frequency
$\omega_{n=1}(F)$ vanishes.
Then the system reaches a 
bifurcation point, the well known Euler instability, beyond which
$\phi_1 \sim \sin{(\pi s/L)}$ becomes the new stable solution 
of the static problem.
For clamped boundary conditions with finite bending moments at $s=0,L$
the critical force is four times larger,
and the shape of the stable solution in the static problem
for $F>F_c$ has the form $\sin^2{(\pi s/L)}$.
Near criticality, the frequencies of higher modes $n=2,3,\ldots$ remain finite.
The dynamics at low frequencies is thus determined
by the fundamental mode alone.
The nonlinear field theory eq.(~\ref{eq:Lagrangian}) may be quantized in the
standard manner by requiring canonical commutation
relations $[\hat\phi(s,t), \hat\Pi(s',t)]=i\hbar\delta(s-s')$ between the field
$\phi$ and its canonically conjugate momentum $\Pi=\sigma\dot\phi$ 
at equal times. In the linear regime, the problem is reduced to 
an infinite number of harmonic oscillators. Introducing oscillator 
lengths $l_k^2=\hbar/(m_k \omega_k)$
with $k=n\pi/L$ and $n=1,2,\ldots$ ,
the amplitudes 
\begin{equation}
\label{eq:quant}
{\mathcal{A}}_k=\frac{l_k}{\sqrt{2}}(a^\dagger_k+a_k)
\end{equation}
are expressed by the standard  
creation and annihilation operators $a^\dagger_k$ and $a_k$ .
The effective masses $m_k$ arising in $l_k$ 
turn out to be $m_{\rm{eff}} \simeq 3/8$ of the beam mass 
for the fundamental mode but are generally mode dependent for 
clamped boundary conditions.
\section{Thermal vibrations}
In the linearized theory, the mean square displacement of the beam
is trivially calculated from the normal mode expansion eq.(~\ref{eq:eigenm}).
Assuming a thermal occupation of the discrete phonon modes one obtains 
a maximum value at the center of the beam, which for unclamped boundary conditions reads
\begin{equation}
\label{eq:x2beam}
\sigma^2=\langle \phi^2(L/2) \rangle=
\frac{l_0^2}{2} \sum_{n\;\rm{odd}} \frac{1}{n\sqrt{n^2-1+\delta}}
\coth{\left( \frac{T_0}{2T}n\sqrt{n^2-1+\delta} \right)}. 
\end{equation} 
Here the scale is set by the oscillator length 
$l_0=\left( \hbar/m_{\rm{eff}} \omega_0 \right)^{\frac{1}{2}}$ and the temperature
$T_0=\hbar\omega_0/ k_{\rm{B}}$ is
associated with the frequency scale of the fundamental mode 
$\omega_0=\omega_1(F\to F_c) \delta^{-\frac{1}{2}}\simeq 1.02\,\omega_1(F=0)$.
The parameter $\delta=(F_c-F)/F_c$ determines the dimensionless distance from 
the critical compression force.
\begin{figure}
\onefigure [width=0.45\linewidth]{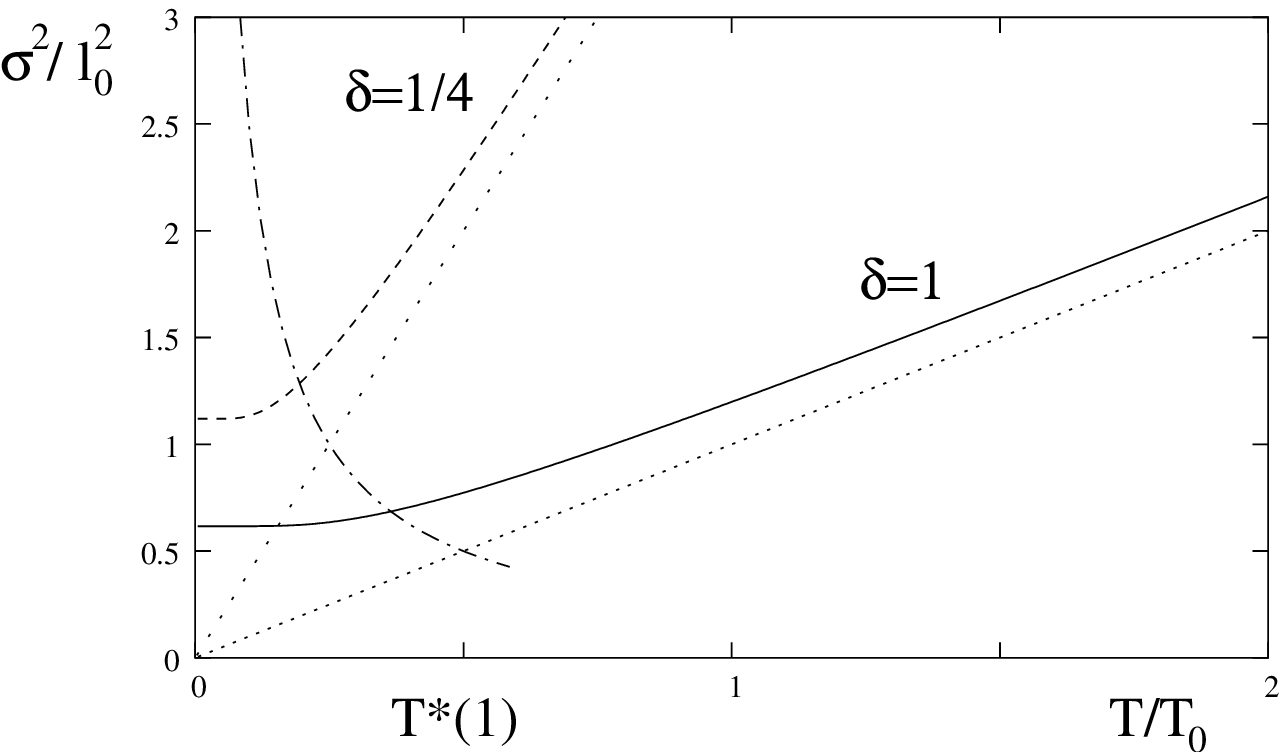}
\caption{Mean square displacement at the center of a clamped beam in linearized theory
for the free ($\delta=1$) and compressed ($\delta=1/4$) case. The dotted
lines show the classical behaviour if one considers only the leading
contribution of the first mode, the dashed-dotted line indicates the
increasing zero point fluctuations versus decreasing crossover-temperature, 
$\{\sigma^2(\delta,T\to 0),T^\star(\delta)\}$, 
for compression approaching the critical force.}
\label{fig:phisq}
\end{figure}
At temperatures larger than $T_0$ one obtains the usual 
equipartition theorem result,
where $\sigma^2 \sim T/\omega_0^2$ increases linearly with $T$
as observed on sufficiently long nanotubes~\cite{Treacy,Schoenenberger}.
For low temperatures the mean square displacement remains finite due to 
zero point fluctuations. As shown in Fig.~\ref{fig:phisq} the crossover to this 
regime occurs at $T^\star \simeq 0.4 T_0$ in the absence of an external force
$\delta=1$, giving accessible temperatures of around $30 {\rm{mK}}$
for typical SWNT's (see Table~\ref{tab:values}). Unfortunately, 
the associated transverse deflection amplitude $l_0$ is only of order
$10^{-2}{\rm{nm}}$ and thus 
beyond accessibility of standard displacement detection techniques.
There are a number of ways, however, to measure such tiny amplitudes,
for instance  by capacitively coupling the beam to the gate of a 
single electron transistor \cite{Knobel} or using its electrostatic interaction  
with a free-standing quantum dot~\cite{Hoehberger}.
Of course the fluctuations are strongly enhanced near 
the critical force $F_c$,
where only the fundamental mode $n=1$ contributes and thus 
$\sigma^2(T=0)$ increases like $l_0^2/\sqrt{\delta}$.
In this case, the problem of measuring the thermal to quantum crossover
is somehow inverted, since 
now the decreasing crossover temperature
$T^\star(\delta) \sim T_0 \sqrt{\delta}$ 
poses a limiting factor for an observation.
Moreover the divergence in eq.(\ref{eq:x2beam}) for $\delta \to 0$
marks the breakdown of the linearized theory.  
To explore the quantitative
enhancement of the fluctuations and the
change of the relevant length and energy scales
near criticality, one has to include the nonlinear terms in 
the bending energy eq.(~\ref{eq:Lagrangian}). 

\section{Buckling instability}

\begin{figure}
\twofigures[width=0.45\linewidth]{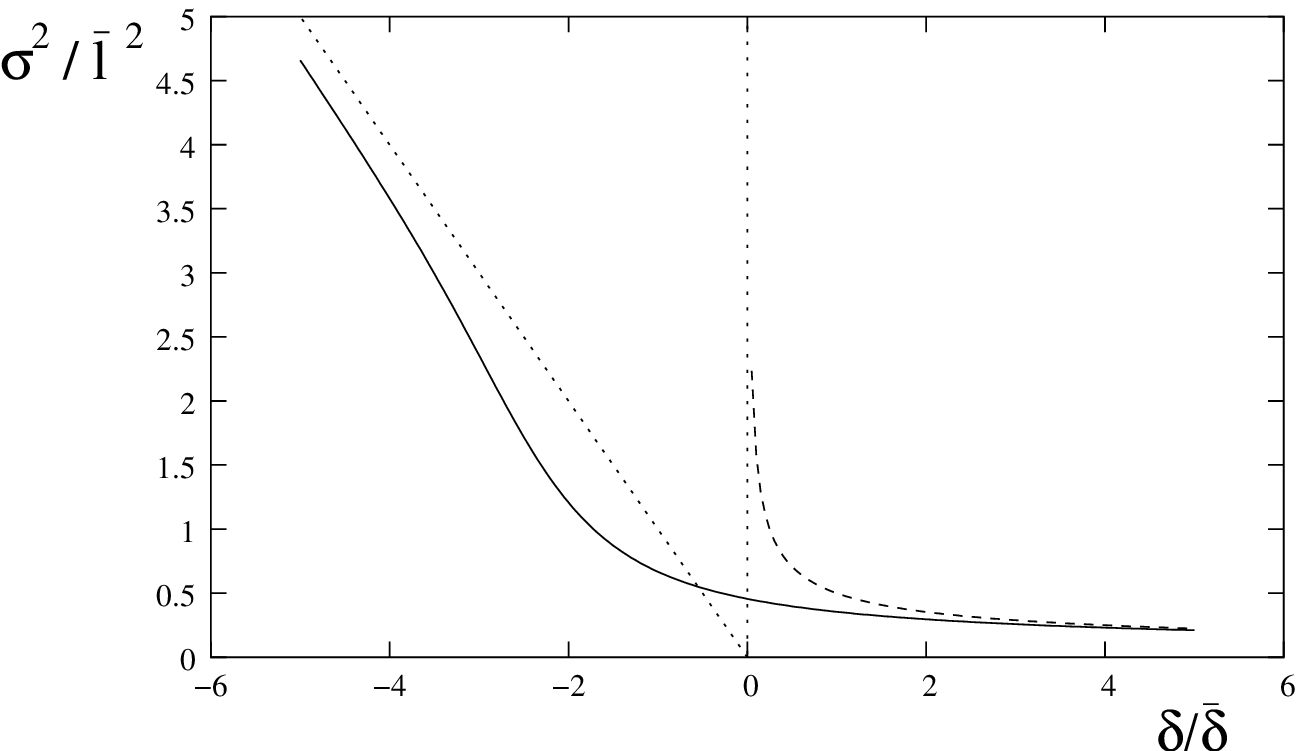}{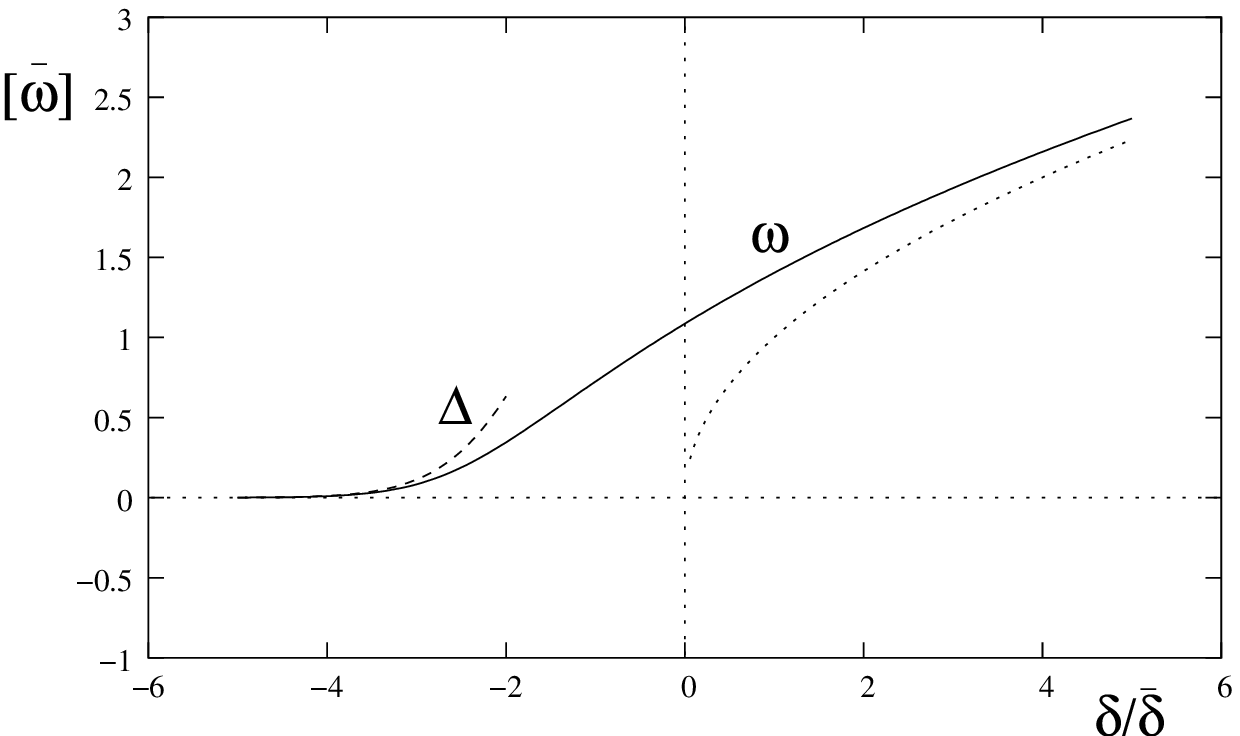}
\caption{Meansquare displacement at zero temperature
at the beams center near the Euler instability.
The solid line is the numerically exact $\sigma^2$ with mean zero,
while the dashed line shows the harmonic approximation 
for $F<F_c$ and the dotted line the result $\sigma^2 \sim y_{\rm{min}}^2$ 
corresponding to the new stable minimum in the broken symmetry phase $F>F_c$.}
\label{fig:y2happrox}
\caption{Numerically exact excitation frequency (solid line)
and tunnelsplitting $\Delta$ (dashed line) 
according to eq.(~\ref{eq:excappr})
in units of $\bar{\omega}$
near the buckling instability. 
The dotted line corresponds to the harmonic approximation for $F<F_c$.}
\label{fig:excenergy}
\end{figure}

In order to describe the behaviour near the buckling instability,
we insert the Fourier expansion eq.(~\ref{eq:eigenm}) 
into the corresponding nonlinear Hamiltonian.
Keeping only the fundamental mode,
since all higher modes have no influence
near criticality
due to their nonvanishing frequencies,
the interacting field theory is reduced to a one particle
problem in terms of the coordinate $\mathcal{A}_1$,
and its canonically conjugate momentum 
${\mathcal{P}} \equiv -i\hbar\partial/\partial{\mathcal{A}_1}$.
The force term generates a negative contribution to the quartic term
in $\mathcal{A}_1$, driving the system unstable.
The nonlinearity in the curvature term, however,
over-compensates this and guarantees stability
even for fixed length of the beam~\cite{Poston}.
For clamped boundary conditions one can 
use the approximate shape $\sin^2{(\pi s/L)}$
which becomes exact near $\delta \to 0$.
One ends up with a quantum mechanical one particle Hamiltonian with
an anharmonic oscillator potential
\begin{equation}
\label{eq:HO4}
H=\frac{{\mathcal{P}}^2}{2m_{\rm eff}}+\frac{\delta}{2}m_{\rm eff} 
\omega_0^2 {\mathcal{A}_1^2} +\frac{b_4}{4}{\mathcal{A}_1^4}
\end{equation}
and an anharmonic coefficient $b_4=(\pi/L)^4 F_cL$.
A similar effective description of quantum effects near the buckling instability
has been derived in~\cite{Wybourne}. The nonlinearity
there, however,  arises from longitudinal stretching while we keep 
the length of the nanotube fixed. 
It is now convenient to define a dimensionless coordinate $y$ by
${\mathcal{A}_1} = \bar{l} y$, where
$\bar{l}=l_0 \,(2\pi^2)^{-1/6} \left(L/l_0 \right)^{1/3}$
is the characteristic magnitude of the deflection,
where the quartic term due to the nonlinear bending energy
is of the same order than the kinetic energy.
The Hamiltonian is thus transformed to a dimensionless form
\begin{equation}
\label{eq:Hdimless}
H=\hbar \bar{\omega} \left(
-\frac{1}{2}\nabla^2+\frac{1}{2}\frac{\delta}{\bar{\delta}} 
y^2 +\frac{1}{4}y^4 \right).
\end{equation}
with $\bar{\omega}=\hbar/(m_{\rm{eff}}\bar{l}^2)$ 
as the characteristic frequency scale
near the critical compression force $F_c$. 
It differs from the fundamental frequency
$\omega_0$ of the classical transverse vibrations by a factor
\begin{equation}
\label{eq:deltaeps}
\frac{\bar{\omega}}{\omega_0} = 
(2\pi^2)^{\frac{1}{3}} \left(\frac{l_0}{L}\right)^{\frac{2}{3}} \equiv \sqrt{\bar{\delta}}
\end{equation}
which also determines the size $\bar{\delta}$ of the critical regime.
For SWNTs with length $L=0.1\rm{\mu m}$, $\bar{\delta}^{1/2}$ is of order 
or smaller than $10^{-2}$ (see Table~\ref{tab:values}).
The potential energy in eq.(~\ref{eq:Hdimless}) exhibits the standard 
Landau bifurcation from a single to a double well as the external
force is increased through its critical value $F_c$.
Indeed our zero dimensional quantum problem  
is equivalent to a one dimensional classical 
Ginzburg Landau theory~\cite{Scalapino}.
Let us consider first the mean square displacement
at the center of the beam. In the harmonic approximation this diverges as
$F$ approaches the critical value from below.
For $F$ much larger than $F_c$ it is simply determined by the stable minimum 
of the effective Landau energy at 
$y_{\rm{min}}=\pm \left( |\delta|/\bar{\delta} \right)^{1/2}$
giving $\sigma^2 =\bar{l}^2 |\delta|/\bar{\delta} $.
As shown in Fig.~\ref{fig:y2happrox},
the exact result smoothly interpolates between those 
two limits giving a finite value
$\sigma(F_c)=0.68\, \bar{l}$ at $F_c$,
which is of order $0.1{\rm{nm}}$ for typical SWNT's (see Table~\ref{tab:values}).
A similar behaviour is found for the lowest excitation frequency 
of the beam. In the harmonic approximation it vanishes like
$\omega_1(F)=\bar{\omega} \cdot \left(\delta/\bar{\delta} \right)^{1/2}$.
Above the critical force the lowest
excitation is the small oscillation in one of the degenerate minima of the 
anharmonic oscillator.
This is true, however, only in a classical description.
Quantum mechanically, the lowest excitation 
is the exponentially small tunnelsplitting $\Delta$
which lifts the degeneracy between the two states localized
in the left or right well of the effective potential.
Again, the exact numerical result for
$\omega=(E_1-E_0)/\hbar$ starts to deviate from the 
harmonic expression at around $\delta \simeq 5\,\bar{\delta}$
and approaches a finite excitation frequency
$\omega=1.1 \bar{\omega}$ at $F_c$ (see Fig.~\ref{fig:excenergy}).
For $\delta < -3\,\bar{\delta}$ it vanishes
exponentially in good agreement with
the WKB result eq.(~\ref{eq:excappr}) for the tunnelsplitting $\Delta$.
It is remarkable that the excitation frequency precisely at $F_c$,
which is of order $2\pi \cdot 0.01 {\rm{GHz}}$
for the parameters of Table~\ref{tab:values},
is no longer related to the characteristic frequency $\omega_0$
of the classical problem but scales like 
$\bar{\omega}=\omega_0(4\hbar\omega_0/F_cL)^{1/3}$,
remaining finite only through a genuine quantum effect.
Unfortunately, the smallness of the size $\bar{\delta}\approx 10^{-4}$
of the critical regime requires fine tuning the compression force F very close
($\delta \simeq \bar{\delta}$) to its critical value in order to 
see deviations from classical behaviour near the buckling instability.

\section{Macroscopic Quantum Coherence (MQC)}

In the regime beyond the critical buckling force, the state of lowest energy
corresponds to a stationary finite deflection amplitude 
$y_{\rm{min}}=\pm \left(|\delta|/\bar{\delta}\right)^{1/2}$ which is lower in energy by
$\hbar\bar{\omega}(\delta/\bar{\delta})^2/4$ than the 
configuration with no deflection.  The direction in which the
buckling occurs is arbitrary, however, in a realistic setup
like that in ref.~\cite{Schoenenberger} the boundary
conditions in the buckled state are likely to break the 
perfect rotation symmetry assumed 
in~\cite{Lawrence}. In this situation, the
transverse deflection is described by a single degree of freedom
with only two degenerate states.   
Quantum mechanically, these states are 
split into a narrow doublet with energy separation $\hbar\Delta$
due to tunnelling.
Sufficiently far above the critical force
its magnitude may be determined from a WKB-calculation in the anharmonic oscillator
potential, giving
\begin{equation}
\label{eq:excappr} 
\Delta = \bar{\omega} \cdot A 
\left(\frac{|\delta|}{\bar{\delta}}\right)^{\frac{5}{4}}
\exp{\left[ -B
\left(\frac{|\delta|}{\bar{\delta}}\right)^{\frac{3}{2}}\right]}. 
\end{equation}
with $A=3.8$ and $B=0.94$~\cite{Weiss}.
Note that the validity of the WKB approximation is limited to $\delta/\bar{\delta}<-2$,
above which the zero point energy of a harmonic approximation in one of the
energy minima reaches the barrier height.
In practice the perfectly degenerate situation is hardly achieved, introducing
some bias energy~$\varepsilon$ which singles out a preferred ground state
in which the beam is bent either to the left or right.
We have thus an effective two state system with Hamiltonian
\begin{equation}
\label{eq:Htwostatesystyem}
\hat{H}=-\frac{\hbar\Delta}{2}\hat{\sigma}_x-\frac{\varepsilon}{2}\hat{\sigma}_z 
\end{equation}
and eigenenergies separated by $\sqrt{(\hbar\Delta)^2+\varepsilon^2}$,
($\hat{\sigma}_x$ and $\hat{\sigma}_z$ are the Pauli spin matrices).
In the absence of any asymmetry~$\varepsilon$
its eigenstates are coherent superpositions $\sim |L\rangle \pm |R\rangle$
of the $\hat{\sigma}_z$ eigenstates $|L\rangle$ and $|R\rangle$,
in which the nanotube is bent towards the left or right. 
The two states $|L\rangle$ or $|R\rangle$
are clearly macroscopically distinct~\cite{Leggett}.
Similar to experiments on flux qubits in SQUID rings
~\cite{vanderWal,Lukens}, the existence of
linear superpositions of these states may indirectly be verified by 
observing the avoided level crossing at~$\varepsilon=0$.
Such a macroscopic quantum coherence experiment with SWNT's requires 
that the small level splitting $\Delta$
can be detected against noise and damping in a mechanical
resonance experiment and moreover, that the asymmetry~$\varepsilon$
can be tuned through zero from any accidental nonzero value
by external means. As shown 
in ~\cite{Poncharal},
spectroscopy of the transverse vibrations of nanotubes
is in principle possible by applying {\em{dc}}- 
plus {\em{ac}}-voltages on a charged nanotube. Moreover
with a capacitive coupling the bias energy $\varepsilon$ 
may be changed via an appropriate electrostatic gate potential.
Due to the still large mass involved, the tunnel splitting is
rather small (around $\Delta=2\pi\cdot 1 \rm{MHz}$ for 
$\delta/\bar{\delta}=-3$), 
and thus coherent superpositions
with an accessible value of $\Delta$ require 
nanotubes close to the buckling instability. 
\begin{table}
\caption{
Characteristic parameters for quantum effects in SWNT's
of length $L=0.1{\rm {\mu m}}$ and diameter $D=1.4{\rm {nm}}$.
We assume a Young's modulus $E=1{\rm{TPa}}$ and an effective wall
thickness $d=5\cdot 10^{-2}{\rm {nm}}$. These parameters are consistent
with recent measurements~\cite{Schoenenberger}.}
\label{tab:values}
\begin{center}
\begin{tabular}{r|r|r|r|r|r|r|r}
$F_c[\rm{nN}]$&$m_{\rm{eff}}[\rm{kg}]$&$\omega_0/2\pi[\rm{GHz}]$&
$l_0[\rm{nm}]$&$T_0[\rm{mK}]$&
$\bar{\delta}^{1/2}$&$\bar{\omega}/2\pi[\rm{MHz}]$&$\bar{l}[\rm{nm}]$  \\ \hline 
0.19 & $1.3\cdot 10^{-22}$ & 1.4 &0.01&
65&$0.6\cdot 10^{-2}$&8.5&0.13
\end{tabular}
\end{center}
\end{table}
Regarding the influence of damping effects, it is known~\cite{Zwerger} 
that the dynamics of a two-level system subject to an ohmic 
dissipation mechanism is determined by the size of the parameter
$\alpha= \eta q_0^2 / (2\pi\hbar)$.
Here $q_0=2 y_{\rm{min}} \bar{l}$ is the distance between the two minima 
and $\eta$ the phenomenological damping parameter which 
appears in the equation of motion
\begin{equation}
\label{eq:etadef}
m_{\rm eff} \ddot{\mathcal{A}}_1 + \eta \dot{\mathcal{A}}_1 + 
\frac{\partial}{\partial \mathcal{A}_1}V(\mathcal{A}_1)=0.
\end{equation}
Coherence in the two level system is present only for $\alpha < \frac{1}{2}$
at $T=0$ and $k_BT<\hbar\Delta/\alpha$ for finite temperature
and very small $\alpha$. 
This requires that the quality factor of the SWNT 
in the uncompressed case (which is related to $\eta$ by $Q=m_{\rm{eff}}\omega_0/ \eta $)
obeys
$Q > 4 |\delta|/ (\pi \bar{\delta}^{3/2})$. For the above value of $\bar\delta$,
this leads to $Q>220$ in the relevant regime $|\delta|\simeq \bar{\delta} $.
This condition does not seem too stringent for SWNTs, note that a
quality factor of $Q=500$ has been reached for Si-based resonators 
in the GHz regime~\cite{Knobel}.

\section{Conclusions}

We have discussed quantum effects in 
the mechanical properties of single wall carbon nanotubes,
in particular zero point fluctuations in the transverse vibrations and 
the possibility to see the analog of MQC in nanobeams below the Euler buckling
instability.
While thermal vibrations of clamped SWNT's down to lengths
$L=0.5{\rm{\mu m}}$ have indeed been observed very recently~\cite{Schoenenberger},
it remains a considerable challenge to measure the tiny zero point
vibrations of order 0.1 nm predicted for SWNTs of length $L=0.1 {\rm{\mu m}}$
near the buckling instability. With the sensitivity attained very
recently with Si-based resonators~\cite{Knobel}, however, reaching this goal 
in the near future seems to be quite realistic. As regards the possibility to see the
analogue of MQC near or below the Euler buckling instability, this  requires to 
tune these systems rather closely below the instability point
and doing spectroscopy with both $dc$- and $ac$-driving.
Provided the methods used for multiwall nanotubes with lengths
of several ${\rm{\mu m}}$~\cite{Poncharal} can be scaled down to clamped
single wall nanotubes, quantum mechanics in its literal meaning
would finally be of relevance in truly mechanical devices.

\acknowledgments
We gratefully acknowledge support 
by the German-Israeli DIP project 
'Coherence, Disorder and Interactions in Coupled Mesoscopic Systems'
and useful discussions with R. Blick, B. Lorentz and C. Sch\"onenberger
on the experimental aspects of our work.

\end{document}